\documentclass[aps,twocolumn,pra,groupedaddress,superscriptaddress,nofootinbib]{revtex4-2}

\usepackage{amsmath,amssymb,amsbsy,subfigure,txfonts}
\usepackage{bbm,times}
\usepackage[T1]{fontenc}
\usepackage{braket}
\usepackage{epsfig} 
\usepackage{color}
\usepackage{graphicx}
\usepackage{dcolumn}
\usepackage{bm}

\usepackage{physics}
\usepackage{xcolor}

\usepackage{hyperref}
\usepackage{cleveref}
\hypersetup{
  colorlinks=true,
  linkcolor=blue,
}
\newcommand{\ave}[1]{\langle#1\rangle}

\providecommand{\openone}{\leavevmode\hbox{\small1\kern-3.8pt\normalsize1}}

\usepackage{soul}
\usepackage{cancel}
\usepackage{float}
\usepackage{multirow}
\newcolumntype{?}{!{\vrule width 1.8pt}}

\usepackage{colortbl}

\graphicspath{{figure/}}

\newcommand*{\opa}{\hat{a}}
\newcommand*{\opadag}{\hat{a}^\dagger}

\newcommand{\opU}[1][k]{\hat{U}_{#1}}
\newcommand{\ops}[2]{\hat{\sigma}_#1^#2}
\newcommand{\opsup}[1]{\dyad{e}{g#1}}
\newcommand{\opsdwn}[1]{\dyad{g#1}{e}}

\newcommand{\ga}{\gamma_\alpha}
\newcommand{\gb}{\gamma_\beta}

\newcommand{\Obs}[1][]{\hat{O}_{#1}}

\newcommand{\rhost}{\rho^*}

\begin{document}


\title{Heating and cooling processes via phaseonium-driven dynamics of cascade systems}

\author{Federico Amato}
\email{federico.amato01@unipa.it}
\affiliation{
 Dipartimento di Ingegneria, Universit\`a degli Studi di Palermo,\\ \quad \quad Viale delle Scienze, 90128 Palermo, Italy
 }
\author{Claudio Pellitteri}
\affiliation{
 Dipartimento di Fisica e Chimica - Emilio Segr\`e, Universit\`a degli Studi di Palermo,\\ \quad \quad Via Archirafi 36, I-90123 Palermo, Italy
}
\author{G. Massimo Palma}
\affiliation{
 Dipartimento di Fisica e Chimica - Emilio Segr\`e, Universit\`a degli Studi di Palermo,\\ \quad \quad Via Archirafi 36, I-90123 Palermo, Italy
}
\affiliation{NEST, Istituto Nanoscienze-CNR, Piazza S. Silvestro 12, 56127 Pisa, Italy
}
\author{Salvatore Lorenzo}
\affiliation{
 Dipartimento di Fisica e Chimica - Emilio Segr\`e, Universit\`a degli Studi di Palermo,\\ \quad \quad Via Archirafi 36, I-90123 Palermo, Italy
}
\author{Rosario Lo Franco}
 \affiliation{
 Dipartimento di Ingegneria, Universit\`a degli Studi di Palermo,\\ \quad \quad Viale delle Scienze, 90128 Palermo, Italy
 }

\date{\today}

\begin{abstract}
The search for strategies to harness the temperature of quantum systems is one of the main goals of quantum thermodynamics. Here we study the dynamics of a system made of a pair of quantum harmonic oscillators, represented by single-mode cavity fields, interacting with a thermally excited beam of phaseonium atoms, which act as ancillas. 
The two cavities are arranged in a cascade configuration, so that the second cavity interacts with phaseonium atoms only after their interaction with the first one. 
We provide exact closed dynamics of the first cavity for arbitrarily long interaction times.
We highlight the role played by the characteristic coherence phase of phaseonium atoms in determining the steady states of the cavity fields as well as that of the ancillas.
Also, we show how the second cavity follows a non-Markovian evolution due to interactions with the ``used'' ancillary atoms, that enable information exchange with the first cavity.
Adjusting the parameters of the phaseonium atoms, we can determine the final stable temperature reached by the cavities. In this way, the cavities can be heated up as well as cooled down. 
These results provide useful insights into the use of different types of ancillas for thermodynamic cycles in cavity QED scenarios.
\end{abstract}

\maketitle


\section{Introduction}
Since the introduction of the \emph{phaseonium} by Scully in \cite{SCULLY1992}, this new ``state of matter'' gave birth to interesting applications, from the early adoptions as a mean for lasing without inversion, refractive index enhancement and correlated spontaneous emission lasers \cite{SCULLY1992_2}, to its later use in a single-heat-bath quantum Carnot engine \cite{SCULLY2003}. Its optical properties were reviewed and developed later on \cite{rathea2021optical, KOZLOV200085}, while applications have been proposed for its optomechanical properties \cite{Nguyen_Soci_Ooi_2016} and for its role as quantum fuel \cite{Turkpens_2016_quantum_fuel}. 
Phaseonium is a three-level lambda system with two almost-degenerate ground states in a coherent superposition.
As simple as it is, the characteristic quantum property of coherence often leads to unexpected and interesting consequences, giving the phaseonium its appeal.

Here, we study the phaseonium as a thermal bath coupled to a multipartite system, within the framework of collision models (CMs) \cite{ciccarello2021, LoFranco_2018}.
As a paradigmatic continuous-variable system, we consider the one composed of two harmonic oscillators, which are physically represented by two single-mode cavities.
We follow phaseonium atoms as they interact with the two subsystems (cavities) one after another, in what is called \emph{cascade} configuration \cite{pichler_quantum_2015, giovannetti_master_2012-2, ramos_quantum_2014, cusumano_interferometric_2018-1, Lorenzo_2015, Lorenzo_PRA, Pellitteri_2023, Milman2005}.  
We give the exact time evolution of the system and see how the atoms of the bath mediate a one-way information flow between the subsystems that lasts until thermalization.
Consequently, while we can trace out the reduced dynamics of the first cavity alone, as it does not ``see'' the second one, the last cavity follows a non-Markovian reduced dynamics.
Although collision models usually end up taking the continuous-time limit of the model, we leave on purpose the description of the dynamics to be \emph{discrete}.

The interaction time between phaseonium and system at discrete time steps plays the role of a control parameter for the overall dynamics, and the effects of gauging it are thus investigated.
Accounting for this, we follow the arguments of Ref.~\cite{Englert_2002} to provide the finite-time quantum map for the system without approximations. As a result, the two cavities reach a thermal state at the same temperature.
We point out that this thermalization process, described by a finite difference master equation, is controlled by the coherence phase of the phaseonium bath atoms.

The system under study in this work represents an interesting open quantum toy model from the point of view of quantum thermodynamics.
In fact, it is \emph{minimal} in showing the properties of \textbf{coherences} in an environment and \textbf{correlations} internal to the system undergoing a transformation.
The exploitation of quantum resources such as coherence is key to reach quantum advantage in thermodynamic tasks, and it is at the heart of Quantum Thermal Engines \cite{vinjanampathy_quantum_2016, bhattacharjee_quantum_2021, oppenheim_thermodynamical_2002, diaz_de_la_cruz_quantum-information_2014} with enhanced efficiency or coefficient of performance \cite{SCULLY2003, Turkpens_2016_quantum_fuel, jaramillo_quantum_2016, dag_multiatom_2016, bengtsson_quantum_2018, levy_quantum_2018}, and fast-charging Quantum Batteries \cite{binder_quantacell_2015,le_spin-chain_2018, ferraro_high-power_2018, shaghaghi_micromasers_2022, shaghaghi_lossy_2023}.
As explained in Ref.~\cite{Latune_2019}, only ``internal coherences'' between degenerate levels of the environment contribute to heat flow.
Therefore, the phaseonium represents the minimum viable system to study the thermodynamic effects of quantum coherences in the environment.
On the other hand, one major challenge in the exploitation of quantum thermodynamic systems like batteries lies in \emph{scalability} \cite{gyhm_quantum_2022, beau_scaling-up_2016, diaz_de_la_cruz_quantum-information_2014}.
The starting point to scalability is the simplest bipartite system in a cascaded configuration.
This is enough to study the effects of intra-system correlations \cite{heat_Lorenzo2015}.
Collectively, we give a description of this minimal system, characterizing the effects of quantum coherences and intra-systems correlations in the thermodynamic context of thermalization.
This will provide the foundations upon which to add complexity, generalizing the results to include more subsystems, adding more complex interactions getting closer to real-world implementations.

The paper is structured as follows. In Section \ref{sec:cms} we present a brief review of the theoretical framework of CMs to tackle open quantum systems' dynamics.
In Section \ref{sec:onesys_description} we describe an optical cavity and the phaseonium atoms with their operators and free-evolution Hamiltonians, as well as their interaction, which constitute the foundation of this work.
In Section \ref{sec:full_time}, we carve out from the cavity-phaseonium interaction the Kraus operators that form the dynamical map for the cavity evolution, giving the stationary conditions and the expressions for long-time stationary states for both cavity and ancillas.
In Section \ref{sec:cascade} we describe the cascade system and its dynamical map with appropriate Kraus operators.
A detailed graphical analysis of the thermalization process of the cavities is given in Section \ref{sec:results}, by following the evolution of their temperatures at each collision step. As a remarkable application, we highlight that the system can be heated up or cooled down to a stable chosen temperature, setting suitable parameters for the incoming phaseonium atoms. To assess the robustness of this process, we also take into account the effects due to stochastic noise in the interaction time or the coherence phase.
Finally, in Section \ref{sec:discussion} we discuss the results in a broader context, from both experimental and theoretical prospects.

\section{Quantum Collision Models}\label{sec:cms}

\begin{figure*}
\includegraphics[width=0.9\textwidth]{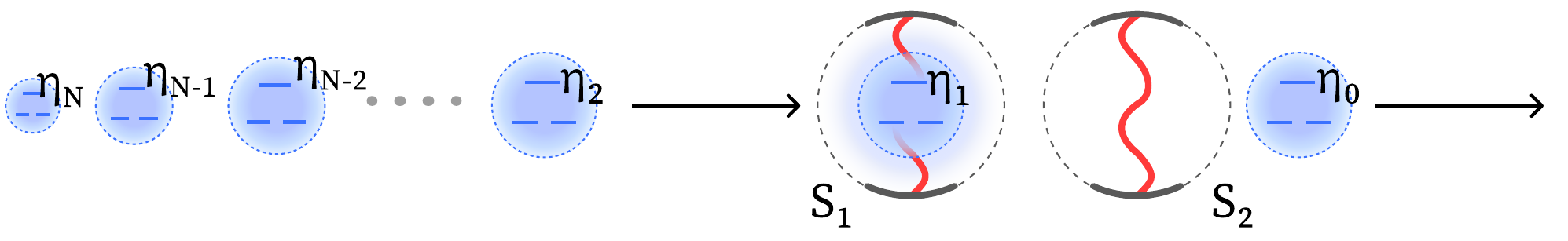}
\caption{\label{fig:phaseonium} Standard collision model for a phaseonium bath interacting with a multipartite system. The system of interest is a cascade of two single-mode cavity fields $S_1$, $S_2$, which is described by a density operator $\rho$. The environment is made up of $N$ three-level atoms in lambda configuration, called phaseonium atoms, all prepared equally. These atoms play the role of ancillas and are described by a density operator $\eta_k$ ($k=0,\ldots, N$). Ancillas travel at speed $v$ and enter the cavities at a rate $r$. They interact with each cavity field for a time $\Delta t$. The speed and rate of phaseonium atoms are selected such that there is at most one ancilla in each cavity at a time.}
\end{figure*}

In this section, we briefly recall the collision model approach \cite{ciccarello2021} to open quantum systems' dynamics \cite{BreuerPetruccione}.
Generally speaking, the dynamics of a quantum system $S$ is said \emph{open} when the system is coupled to a bath $B$ whose time evolution cannot be investigated or is inaccessible.
In the simplest collision model, the bath $B$ is taken as a large discrete collection of elementary systems---called \emph{ancillas}---that interact \emph{one at a time} with the system, where each of these \emph{collisions} is described by unitary two-body interactions. 
Such a collision model decomposes the complex system-bath dynamics into simple contributions. 
An intuitive implementation of collision models is, for example, the \emph{micromaser} \cite{Filipowicz:86-1, Filipowicz:86-2}. 

The initial joint state of the bath $B$, made up of identical ancillas $\eta_k$ ($k=1,2,\dots,N$), and the system $S$ is assumed to be a product state 
$\chi_0=\rho_0\otimes\eta_1\otimes\eta_2\otimes\ldots\otimes\eta_N$.
One by one, ancillas collide with the system for a short time $\Delta t$ at discrete time steps labeled by the same $k$ as that of the colliding ancilla. The dynamics are thus discrete.
Let us denote with $\hat{H}_S$ and $\hat{H}_{\eta}$ the free Hamiltonians of $S$ and the $k$-th ancilla $\eta_k$, respectively, which define the total Hamiltonian $\hat{H}_0 = \hat{H}_S + \hat{H}_\eta$. Indicating the system-ancilla interaction with \(\hat{V}_k\), the dynamics of each $k$-th collision is ruled by the unitary evolution operator
\begin{equation}
    \opU=e^{-i\Delta t(\hat{H}_S+\hat{H}_\eta+\hat{V}_k)}
\,.\end{equation}

Three assumptions are now required to ensure a Markovian behaviour of the model (i.e., governed by a master equation in Lindblad form):
\begin{itemize}
    \item ancillas do not interact with each other;
    \item ancillas are initially uncorrelated;
    \item each ancilla interacts with S only once for a time $\Delta t$.
\end{itemize}

At each step, the whole state evolves according to $\chi_k {=} \opU\chi_{k-1}\opU^\dagger\,$, starting from $\chi_1{=}\opU[1]\chi_0\opU[1]^\dagger$. 
The corresponding state of the system at each step, $\rho_k$, is obtained by tracing out the ancillas' degrees of freedom. Using the partial trace $\Tr_j$ over the degrees of freedom of the $j$-th ancilla, one finds, recursively,
\begin{eqnarray}
    \rho_k &&= \Tr_{\mathcal{B}}\{\chi_k\} \nonumber
    {=}\Tr_k\{\opU[k]\cdots\Tr_1\{\opU[1]\rho_0\otimes\eta_1\opU[1]^\dagger \}\cdots\otimes\eta_k\opU[k]^\dagger\}\,.
\end{eqnarray}
This suggests defining the quantum \emph{collision map} $\mathcal{E}$ as
\begin{equation}\label{def:collision_map}
    \rho_k=\mathcal{E}[\rho_{k-1}] = \Tr_k\{\opU(\rho_{k-1}\otimes\eta_k)\opU^\dagger\}\,,
\end{equation}
which shows that the state of $S$ at step $n$ only depends on that at the previous step $n-1$: the dynamics has \emph{no memory} of its past history, as is expected for Markovian dynamics (see, for instance, Ref.~\cite{nonMarkovian}, for a review on Markovian and non-Markovian processes with memory effects).

The steady-state $\rhost$ of the system is obtained by the \emph{fixed point} of the dynamical map, that is the state such that $\mathcal{E}[\rhost]=\rho^*$. If the fixed point is unique, the collision map is \emph{ergodic}. Moreover, if any initial state $\rho_0$ eventually tends to the same steady state $\rhost$, the map is said to be \emph{mixing} \cite{richter_ergodicity_1996, burgarth_2007, burgarth_2013}.

\section{System and phaseonium bath}\label{sec:onesys_description}

We begin our discussion by introducing the main actors: the open system we want to look at and the environment in which it is positioned. Figure \ref{fig:phaseonium} depicts the model of interest for a cascade two-cavity system interacting with a beam of phaseonium atoms, which will be analysed in Sec. \ref{sec:cascade}. 

Here, we start considering a single-mode optical cavity as system $S$.  
The cavity behaves like a single-mode harmonic oscillator whose Hamiltonian is \cite{BreuerPetruccione}
\begin{equation}\label{eq:system_hamiltonian}
    \hat{H}_S = \hbar\omega_c \left(\opadag \opa + \frac{1}{2}\right)
\;,\end{equation}
where $\opadag$ and $\opa$ are, respectively, the creation and annihilation operators stemming from the canonical quadrature operators position $\hat q=\frac{1}{2}(\opadag+\opa)$ and momentum $\hat p=\frac{i}{2}(\opadag-\opa)$, satisfying the usual commutation relation $\comm{\hat p}{\hat q} = i$.

The cavity is coupled to the environment via short-time interactions with ancilla systems pumped in the cavity itself. 
Every ancilla $\eta_k$ is a three-level lambda system. Its states are denoted by $\ket{e}$, $\ket{g_1}$, and $\ket{g_2}$, where $\ket{e}$ represents the excited state while $\ket{e}$, $\ket{g_1}$ are two ground states. 
In this basis, a thermal ancilla can be represented by the density operator
\begin{equation}\label{eq:ancilla-thermal}
    \eta_{th} = 
\begin{pmatrix}
\begin{array}{ccc}
    \alpha^2 & 0 & 0 \\
    0 & \frac{1}{2} \beta^2 & 0 \\
    0 & 0 & \frac{1}{2} \beta^2 \\
\end{array}
\end{pmatrix}
\;,\end{equation}
with the condition $\abs{\alpha}^2~=~1-\abs{\beta}^2$ to have a unitary trace.

From such a thermal state, one can create coherences between ground levels, obtaining the coherent ancilla state which defines the so-called \emph{phaseonium} \cite{SCULLY1997}.
Note that a little energy shift $\epsilon$ is created in the process.
As seen in \cite{SCULLY2002}, we can assume that this shift is negligible and the ground-states energy remains degenerate, writing the phaseonium density matrix as
\begin{equation}
\label{def:ancilla-state}
    \eta_k = 
\begin{pmatrix}
\begin{array}{ccc}
    \alpha^2 & 0 & 0
\\
    0 & \frac{\beta^2}{2} + \epsilon & \frac{\beta^2}{2}  e^{-i \phi } 
\\[.3em]
    0 & \frac{\beta^2}{2}  e^{i \phi } & \frac{\beta^2}{2}-\epsilon \\
\end{array}
\end{pmatrix} \approx 
\begin{pmatrix}
\begin{array}{ccc}
    \alpha^2 & 0 & 0
\\
    0 & \frac{\beta^2}{2} & \frac{\beta^2}{2} e^{-i \phi} 
\\[.3em]
    0 & \frac{\beta^2}{2}  e^{i \phi } & \frac{\beta^2}{2} \\
\end{array}
\end{pmatrix}
\,.\end{equation}
 
One can thus write a simple free Hamiltonian $H_\eta$ for the phaseonium as
\begin{equation}
    \hat{H}_\eta = \frac{\hbar\omega_\eta}{2}(\ops{1}{+}\ops{1}{-} + \ops{2}{+}\ops{2}{-})
\;,\end{equation}
with ladder operators $\hat\sigma_i^\pm$ acting on the different ground levels of the ancilla
\begin{eqnarray}\label{ladderop}
    \ops{1}{+} &= \opsup{_1} \: , \quad  \ops{1}{-} = \opsdwn{_1} \,, \nonumber\\
    \ops{2}{+} &= \opsup{_2} \: , \quad  \ops{2}{-} = \opsdwn{_2}
\;.\end{eqnarray}
Notice that the operators $\ops{i}{\pm}$ ($i=1,2$) refer to the ground state $\ket{g_i}$.
We choose a resonant coupling with $\omega_c = \omega_\eta \equiv \omega$ and use the interaction picture to leave the free evolution of both cavity and bath out of the analysis.
So, indicating with $\Omega$ the coupling strength, the total system-environment Hamiltonian at the $k$-th collision is given by the interaction term
\begin{equation}\label{def:interaction}
    \hat{V}_k = \hbar\Omega\left[ \opa(\ops{1}{+} + \ops{2}{+}) + \opadag(\ops{1}{-} + \ops{2}{-}) \right] \,.
\end{equation}
\\

\section{Cavity-phaseonium evolution}\label{sec:full_time}

The time-evolution operator $\hat{U}_k = \exp(i\hat{V}_k\Delta t)$, with an arbitrary-long interaction time $\Delta t$, can be written in the basis of ancilla states $\ket{e}$, $\ket{g_1}$, $\ket{g_2}$, as (see Appendix~\ref{apx:time-evolution-expansion})
\begin{equation}\label{def:time-evolution-matrix}
    e^{-i\theta V_k} =
    \begin{pmatrix}
        \hat{C} & -i S^\dagger & -i S^\dagger \\
        -i \hat{S} &  \tfrac{1}{2}(\hat{C}' +\mathbb{I}) & \tfrac{1}{2}(\hat{C}' -\mathbb{I}) \\
        -i \hat{S} &  \tfrac{1}{2}(\hat{C}' -\mathbb{I}) & \tfrac{1}{2}(\hat{C}' +\mathbb{I})
    \end{pmatrix}
\;,\end{equation}\\
where $\theta$ is the accumulated Rabi phase $\hbar\Omega\Delta t$, while $\hat{C}$, $\hat{C}'$ and $\hat{S}$ are the photonic operators
\begin{subequations}
\label{def:photonic-ops}
\begin{align}
    \hat{C} &= \cos(\theta\sqrt{2\opa\opadag}), \\[.5em]
    \hat{C}' &= \cos(\theta\sqrt{2\opadag\opa}), \\
    \hat{S} &= \opadag\frac{\sin(\theta\sqrt{2\opa\opadag})}{\sqrt{2\opa\opadag}} 
\;.\end{align}
\end{subequations}

Thanks to this representation, we are now able to write the map acting on the cavity at each step in its Kraus decomposition, which is
\begin{equation}
    \rho_{k+1} = \mathcal{E}\left[\rho_{k}\right] =
    \Tr_k\left\{e^{-i\theta V_{k+1}}\,\rho_k\eta_{k+1}\,e^{i\theta V_{k+1}}\right\} 
    = \sum_{i=0}^{4} \hat E_i \rho_k \hat E_i^\dagger, \label{eq:kraus_map1}
\end{equation}
with the operators $\hat E_i$ given by
\begin{subequations}
\label{def:kraus1}
\begin{align}
    &&\hat{E}_0 = \sqrt{1 - \frac{\ga}{2} - \frac{\gb}{2}}\, \mathbb{I} \;,\; 
    \hat{E}_1 = \sqrt{\frac{\gamma_\alpha}{2}}\,\hat C,\\
    &&\hat{E}_2 = \sqrt{\gamma_\alpha}\, \hat S \;,\; 
    \hat{E}_3 = \sqrt{\frac{\gamma_\beta}{2}}\,\hat C'\;,\; 
    \hat{E}_4 = \sqrt{\gamma_\beta}\,\hat S^\dagger ,
\end{align}
\end{subequations}
where $\ga=2\alpha^2$ and $\gb=\beta^2(1+\cos\phi)$. It is important to observe that the Kraus operators specified in Eq.~(\ref{def:kraus1}) have a specific structure that precludes the generation of coherences. Exploiting the relations $\hat{C}\hat{C}{+}2\hat{S}^\dagger\hat{S}=\hat{C}'\hat{C}'{+}2\hat{S}\hat{S}^\dagger =\mathbb{I}$, it is possible to write a finite difference master equation for the cavity alone at each interaction step, as
\begin{equation}\label{eq:finitediff_ME}
    \Delta\rho_k=\sum_{i=0}^{4} \hat E_i \rho_k \hat E_i^\dagger-\rho_k=\sum_{i=1}^{4} \mathcal{D}[\hat E_i]\rho_k
\;,\end{equation}
in which we adopt the usual definition for dissipator $\mathcal{D}[\hat o]\rho{=}\hat o\rho\hat o^\dagger{-}1/2\{\hat o^\dagger\hat o,\rho\}$.  


\subsection{Cavity Steady State}\label{sec:onesys_steadystate}
Thanks to Eq.~(\ref{eq:finitediff_ME}), the stationary state $\rhost$ of the system can be found solving the equation
\begin{equation}\label{eq:steady-state-hypothesis}
   \Delta\rhost=0 \,.
\end{equation}
The solution of the above equation (see Appendix~\ref{apx:steady-state} for details) can be expressed by the diagonal density matrix
\begin{equation}\label{def:steady-state}
    \rhost = \sum_n^\infty\frac{\left( \gamma_\alpha / \gamma_\beta \right)^n}{Z}\dyad{n} \,, 
\end{equation}
where $1/Z=\rhost_{00}$ is the first element of the density matrix.
We can then make a parallel with a standard Gibbs state \cite{gogolin2016} and set $(\gamma_\alpha / \gamma_\beta)^n = \exp(-n\hbar\omega / K_BT_\phi)$, where $K_B$ is the Boltzmann constant, which defines an effective temperature $T_\phi$
\begin{equation}
    \exp{\left(-\frac{\hbar\omega}{K_B T_\phi}\right)} 
    = \frac{\gamma_\alpha}{\gamma_\beta} 
    \;\; \Rightarrow \;\;
    K_BT_\phi = -\hbar\omega/\ln{\left(\frac{\gamma_\alpha}{\gamma_\beta}\right)} \,. \label{def:steady-temperature}
\end{equation}
Since $\gamma_\alpha / \gamma_\beta = 2|\alpha|^2/(|\beta|^2(1+\cos\phi))$, we emphasize that this effective temperature $T_\phi$ fundamentally depends on the coherence phase $\phi$ and on the excited-state population $\alpha$ of the phaseonium.
Thus, \emph{for every initial state} of the cavity in contact with the phaseonium beam, the system will end up in a thermal Gibbs state
\begin{equation}\label{def:gibbs-state}
    \rhost = \sum_n^\infty\frac{\exp(-E_n/K_B T_\phi)}{Z}\dyad{n} \,. 
\end{equation}
The above state is independent of the collision duration $\Delta t$, meaning that this is also the steady state of the continuous-limit master equation, as shown in Appendix~\ref{apx:continuous-time}.

\begin{figure}
    \includegraphics[width=0.48\textwidth]{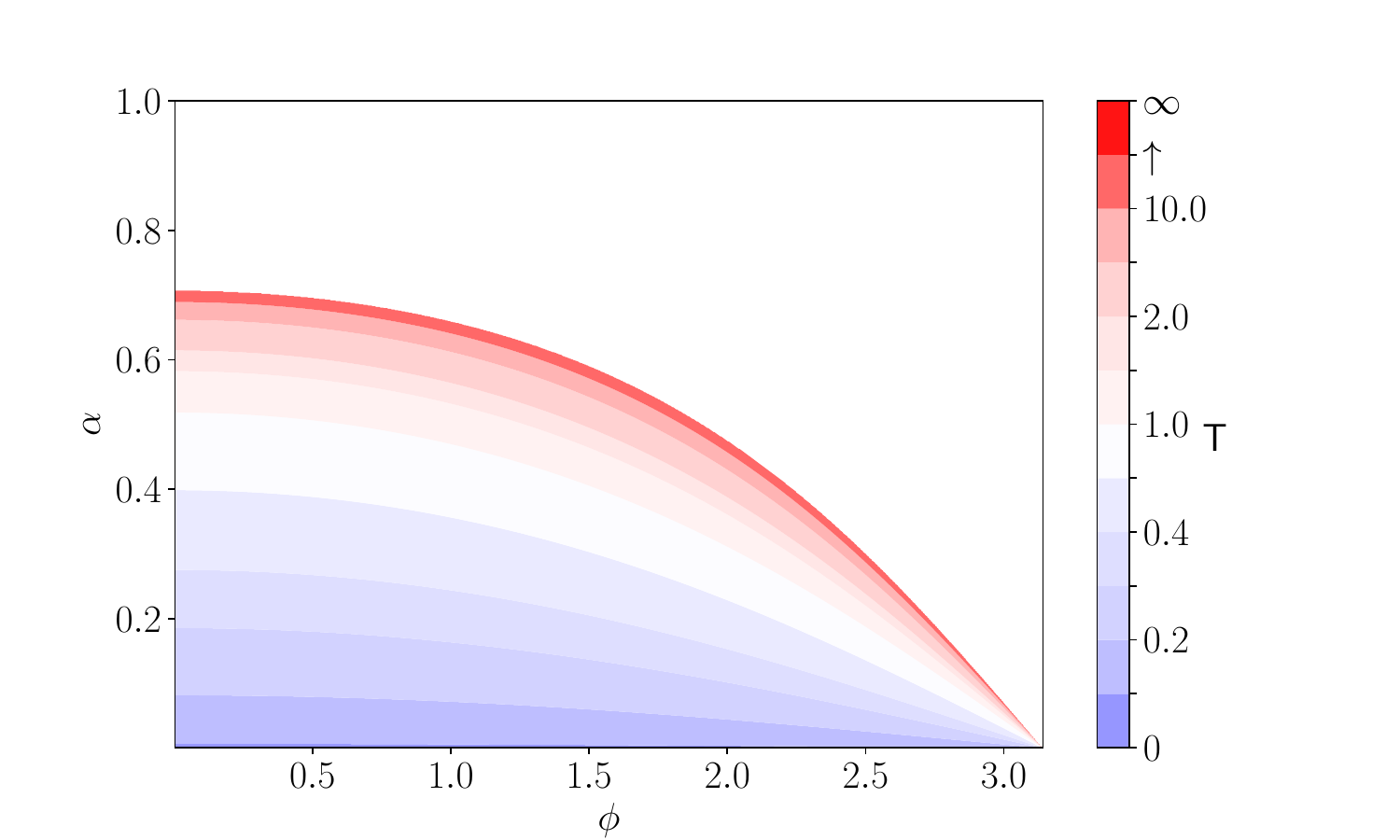}
    \caption{Range of steady-state temperatures of one cavity in contact with a phaseonium bath, as a function of excited-state population $\alpha$ and ground-state coherences $\phi$ of phaseonium atoms. Temperature is symmetric in $\phi$, so it is shown in the range $0 \le \alpha \le 1$,  $0 \le \phi \le \pi$. Temperature is given in units of $K_B/(\hbar\omega)$.}
    \label{fig:steady-state-temperatures}
\end{figure}

The range of temperatures spanned by ancilla parameters $\alpha$ and $\phi$ is shown in Figure~\ref{fig:steady-state-temperatures}.
Seeing that temperature is positive-defined, our Gibbs state representation works as long as we have low-excited ancilla probability such that \(\ga/\gb<1\).
Two special cases emerge: (i) when \(\abs{\alpha}^2=0\), the system depletes itself and ends up in a vacuum thermal state (the atoms only absorb photons from the cavity radiation field); (ii) when $\abs{\alpha}^2{=}\abs{\beta}^2\cos^2(\phi/2)$, that is \(\ga/\gb=1\), Eq.~(\ref{def:steady-temperature}) does not hold, leading to an infinite temperature. 
Thus, for $\ga/\gb\geq 1$ this description is no more accurate and we can conclude that, under these conditions, the hypothesis of Eq.~(\ref{eq:steady-state-hypothesis}) of the existence of a stationary state is no more valid.
Also, the value $\phi = \pi$ is to be excluded, since it makes the ratio $\ga/\gb$ undefined. 

Since we have not specified the initial state of the system, we can infer that this fixed point state \(\rhost\) of the collision map is unique and reached whatever the initial state of the system.

The final thermalization temperature $T_\phi$ of the cavity is in general different from the temperature $T$ of the starting thermal ancillas $\eta_{th}$ defined in Eq.~(\ref{eq:ancilla-thermal}), before the ground-state coherences are created.
Indeed, there is an energetic cost in creating quantum coherences.
This energy difference can be expressed by
\begin{equation}
\frac{\hbar\omega}{K_B T_\phi} - \frac{\hbar\omega}{K_B T} = \ln(1+\cos(\phi))\,.
\end{equation}
Assuming that between the system and the phaseonium atoms there is only heat exchange, it is possible to define an {\it apparent} temperature $\mathcal{T}$ even if the atoms are not in a thermal state~\cite{Latune_2019, Man_2020}. This apparent temperature is defined as 
\begin{equation}\label{def:apparent-ancillas-temperature}
    K_B \mathcal{T}= \hbar\omega\left[
    \ln\left(
        \frac{\Tr_S\left\{(\hat{\sigma}^+_1+\hat{\sigma}^+_2)(\hat{\sigma}^-_1+\hat{\sigma}^-_2)\eta_k\right\}}
        {\Tr_S\left\{(\hat{\sigma}^-_1+\hat{\sigma}^-_2)(\hat{\sigma}^+_1+\hat{\sigma}^+_2)\eta_k\right\}}\right)
    \right]^{-1},
\end{equation}
where $\hat{\sigma}^\pm_{i}$ are the atomic operators of the three-level ancillas defined in Eq.~(\ref{ladderop}).
Remarkably, it can be shown that this coincides with the expression of the stationary effective temperature of the system $T_\phi$ given in Eq.~(\ref{def:steady-temperature}).

\subsection{Ancilla Steady State}
Thanks to the above result, we can also retrieve the ancilla's stationary state once the cavity has thermalized. 
Using Eqs.~({\ref{def:gibbs-state}) and (\ref{def:time-evolution-matrix}), and performing the partial trace over the system degrees of freedom, we obtain
\begin{equation}\label{eq:steady-ancilla}
    \eta^{\text{st}}=\Tr_S\left\{\opU \rhost\eta_k\opU \right\}=
    \begin{pmatrix}
        \abs{\alpha}^2 & 0 & 0 \\[.5em]
        0 & \tfrac{\abs{\beta}^2}{2} &\tfrac{\abs{\beta}^2}{2}\Gamma^\dagger\\[.5em]
        0 & \tfrac{\abs{\beta}^2}{2}\Gamma & \tfrac{\abs{\beta}^2}{2}
    \end{pmatrix},
\end{equation}
with modified coherences given by
\begin{equation}
\Gamma = \cos(\phi)-i \sin(\phi)\Tr{\hat{C}'\rhost}\,.    
\end{equation}

We stress that this evolved ancilla carries the same steady-state system temperature, given only by the real part of coherences, as in Figure~\ref{fig:steady-state-temperatures}.
Ancillas can thus be exploited again to let another cavity thermalize at the same temperature. 
We shall investigate this situation in the following sections.
We explore the possibility to have ancilla decohere due to interactions with the external environment passing from one cavity to the other in Appendix~\ref{apx:decoherence}.

\section{Cascade cavity-phaseonium evolution}\label{sec:cascade}
We now take a system of two identical single-mode cavities, $S_1$, $S_2$, subject to the same collision interaction with a beam of phaseonium atoms (see Figure~\ref{fig:phaseonium}). 
They are arranged in a way that every ancilla atom first interacts with the subsystem $S_1$, modifying it, and successively with the second subsystem $S_2$. 
This is called \emph{cascade quantum system}~\cite{giovannetti_master_2012-2}.

We must consider the total evolution of the bipartite system $S$ made of the two cavities $S_1$ and $S_2$. 
This translates in the subsequent application of unitary interaction operators $\opU[j,k]$ ($j=1, 2$ corresponds to the cavity $S_j$) to the overall density operator of both system and ancilla. Indicating with $\rho_k$ and $\eta_k$, respectively, the density operators of the two-cavity system $S$ and of the ancilla atom at the $k$-th interaction step, extending Eq.~(\ref{def:collision_map}) we have
\begin{equation}\label{eq:cascaded_evolution}
    \rho_k = \mathcal{E}[\rho_{k-1}] = \Tr_k\left\{ \opU[2,k]\opU[1,k]\rho_{k-1}\otimes\eta_k\opU[1,k]^{\dagger}\opU[2,k]^{\dagger}\right\}\,.
\end{equation}
The crucial property of such dynamics relies on the fact that cavity $S_1$ ``does not see'' the following cavity $S_2$, so its evolution is again described by Eq.~(\ref{eq:kraus_map1}). $S_1$ interacts only with ancillas in their initial state $\eta_k$. After that interaction, however, the ancilla state is modified. 
The dynamics of subsystem $S_2$ thus depend on that of subsystem $S_1$. 
By the action of common ancilla atoms, the two subsystems get correlated. 
This is explained by the fact that the common phaseonium ancilla atoms mediate the interaction between the cavity subsystems.
If we trace out $S_1$ from Eq.~(\ref{eq:cascaded_evolution}), we get the state of the cavity $S_2$ alone as
\begin{equation}\label{eq:s2_unitary_evolution}
\begin{split}
    \rho_{2,k} &= \Tr_{S_1}\left\{\Tr_k\left\{ 
        \opU[2,k]\opU[1,k](\rho_{k-1}\otimes\eta_k)\opU[1,k]^\dagger\opU[2,k]^{\dagger}
    \right\}\right\} \\
    &= \Tr_k\left\{ 
        \opU[2,k]\Tr_{S_1}\left\{\opU[1,k](\rho_{k-1}\otimes\eta_k)\opU[1,k]^{\dagger}\right\}\opU[2,k]^{\dagger}\right\}
\;.\end{split}
\end{equation}
Cavity $S_2$ does not follow a Completely Positive Trace Preserving (CPT) map and its dynamics is non-Markovian. 

Following an analogous analysis of that of Section~\ref{sec:full_time}, it can be shown that the map of the two-cavity system $S$ is represented by the Kraus operators
\begin{subequations}\label{def:kraus2}
\begin{eqnarray}
    \hat E_0 &&= \sqrt{1 - \frac{\ga}{2} - \frac{\gb}{2}}\, \mathbb{I}\,, \\[.5em]
    \hat E_1 &&= \sqrt{\frac{\gamma_\alpha}{2}}\,( \hat C{\otimes}\hat C{-2}\hat S{\otimes}\hat S^\dagger)\,, \\[.5em]
    \hat E_2 &&= \sqrt{\gamma_\alpha}\, (\hat S{\otimes} \hat C'{+}\hat C{\otimes} \hat S)\,,\\[.7em]
    \hat E_3 &&= \sqrt{\gamma_\beta}\,(\hat S^\dagger{\otimes}\hat C{+}\hat C'{\otimes} \hat S^\dagger) \,, \\[.5em]
    \hat E_4 &&= \sqrt{\frac{\gamma_\beta}{2}}\,(\hat C'{\otimes} C'{-2}\hat S^\dagger{\otimes} \hat S) \,.
\end{eqnarray}
\end{subequations}
With these operators, we can finally determine the finite difference master equation for the evolution of the two-cavity system $S$
\begin{equation}\label{eq:final-lindblad-meq}
    \Delta\rho_k=\sum_{i=1}^4\mathcal{D}\left[ \hat{E}_i\right]\rho_{k}\,.
\end{equation}

\section{Application}\label{sec:results}

Having found the master equation of Eq.~(\ref{eq:final-lindblad-meq}), we can follow the evolution of the two cavity fields while interacting with the beam of phaseonium atoms.
As we have seen, this is a thermalization process. In this section we provide an application of this controlled process which leads to heating and cooling the cavities.  

Let us summarize in a simple protocol how to exploit the model we developed so far:
\begin{enumerate}
    \item Choose a temperature $T^*$ and pick a pair of parameters $\{\alpha^*,\,\phi^*\}$ from Fig.~\ref{fig:steady-state-temperatures} corresponding to that temperature, or use Eq.~(\ref{def:steady-temperature}) to find them;
    \item Prepare an ensemble of three-level lambda thermal atoms characterized by an excited-state population $\alpha^*$;
    \item Split the degenerate ground doublet of those atoms by creating a certain amount of coherence characterized by the phase $\phi^*$---now you have an ensemble of phaseonium atoms;
    \item Inject a beam of phaseonium atoms one at a time, inside the two cavities, placed in a cascade configuration, so that the atoms interact with each cavity for a time $\Delta t^*$;
    \item After some time, the temperature of each cavity will be the initially chosen temperature $T^*$.
\end{enumerate}
Recall that, in general, this temperature will be \emph{different} from that of the initially prepared thermal ancillas at step 2, due to the energetic cost of creating quantum coherences in the atomic state.  

With this protocol in mind, we show in the following some results about the exploitation of phaseonium atoms to drive the dynamics of the two cavity fields towards desired target states.

\begin{figure}
    \includegraphics[width=0.97\linewidth]{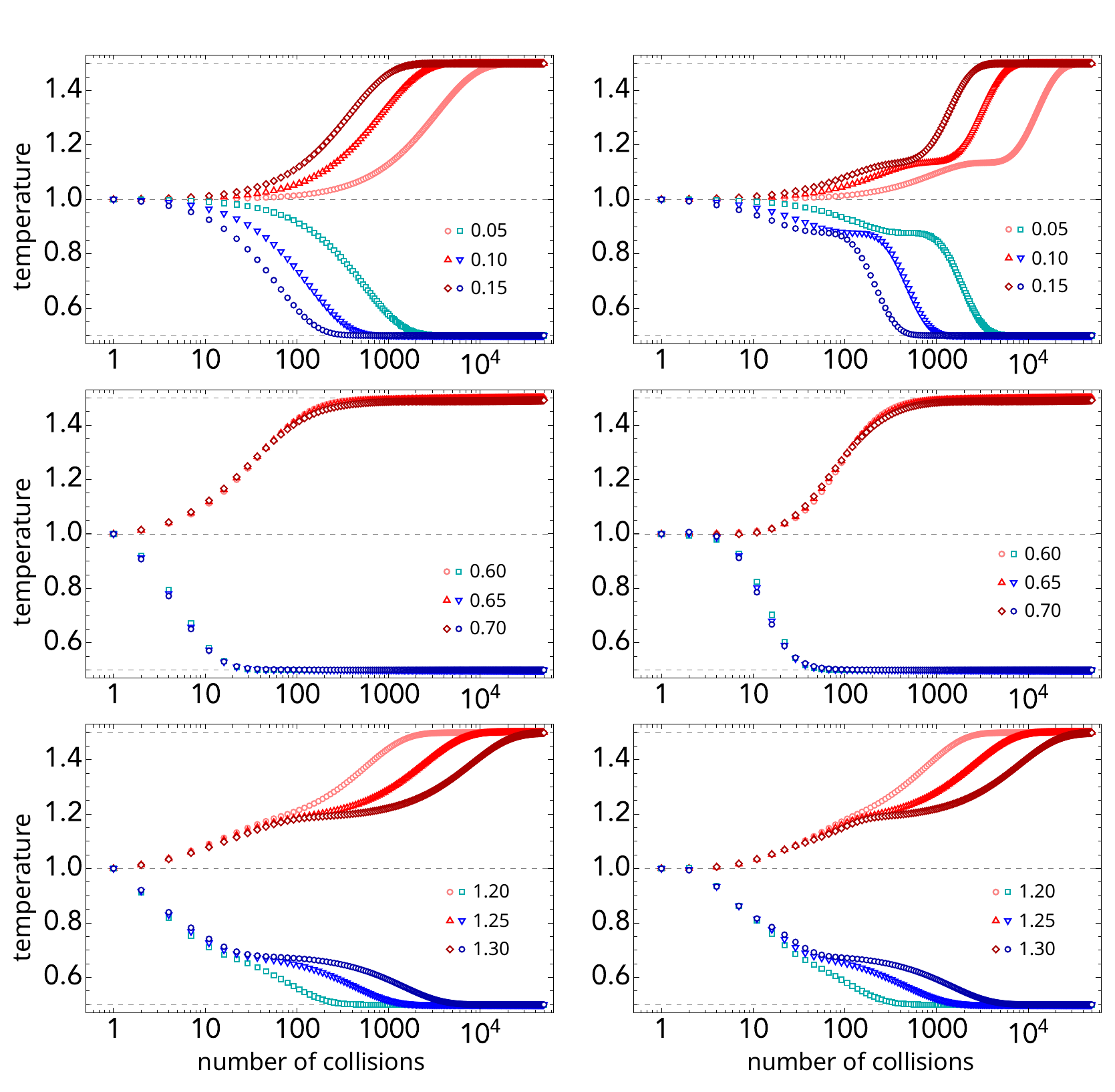}
    \caption{Behaviour of temperature of the first cavity (left column) and second cavity (right column) taking different values for the collision time. The initial state of both the subsystems is taken to be a thermal state at $T=1$. In each plot is shown the thermalization process obtained involving ancillas with a different amount of coherences (but the same $\alpha{=}0.25$). More precisely, for the hot curves $\phi=1.5855$ and for the cold curves $\phi=2.4043$, that correspond to stationary temperatures for the cavities $T=1.5$ and $T=0.5$, respectively. We observe as increasing the collision time the thermalization process becomes faster for small $\Delta t$, does not change for $\Delta t \sim 0.6$ and becomes slower for $\Delta t>1.2$. Interaction times are given in units of $1/\Omega$, while temperature is shown in units of $K_B/(\hbar\omega)$.}
    \label{fig:heatingcooling}
\end{figure}

\subsection{Heating and Cooling the Cavities}
Firstly, we demonstrate how general the protocol is: choosing any temperature $T^*$, one can always find a pair of phaseonium parameters to heat up as well as cool down the cavities towards that temperature.

In Figure~\ref{fig:heatingcooling} we show that for a fixed excited-state population $\alpha$, gauging the coherence phase $\phi$, we can control the evolution of the system to make it get hotter or colder. In particular, it is seen how starting from a thermal state at $T=1.0$ for both cavities, we can appropriately choose the values of the phase $\phi$ such that the two subsystems thermalize either to a temperature $T=1.5$ or to $T=0.5$.
We notice how cavity $S_1$ is always the first to thermalize: this is clear, given the unidirectionality of the cascade two-cavity system. The second cavity $S_2$ eventually reaches the same stable temperature of $S_1$.
Regarding the speed of the thermalization process, it depends on all three parameters, $\alpha$, $\phi$ and $\Delta t$.
Having already prepared our phaseonium atoms, we can then easily change the interaction time to speed up or slow down the process.

As displayed in the three rows of Figure~\ref{fig:heatingcooling}, the behaviour of the thermalization as a function of the number of collisions depends on the atom-cavity interaction time $\Delta t$, but in each row this dependence is different: 
for interaction times of the order of $0.10$ the process speeds up by increasing $\Delta t$ (panels in the top row); for interaction times of the order of $1.25$ the process slows down by increasing $\Delta t$ (panels in the bottom row); for intermediate interaction times the change in speed is negligible (panels in the central row).
This nonlinear behaviour can be understood by looking at the expression of the time evolution of the photon number operator $\Delta \ave{\hat{n}}_k = \ave{\hat{n}}_k - \ave{\hat{n}}_{k-1}$.
The bigger this is, the faster a cavity will gain or lose photons to the ancillas, reaching a steady state faster.
The master equation is given in Eq.~(\ref{eq:n-meq-apx}) of Appendix~\ref{apx:operators-evolution}, where we find the master equation for a generic operator.
Considering this rate of photon gain or photon loss to the ancillas, we see that its magnitude is periodical in $\theta$ via a complex sum of squared sines.
Heuristically, we can think that when an ancilla stays inside the cavity for a long enough time, a photon emitted (absorbed) by the ancilla can also be re-absorbed (emitted back) by the same ancilla to the cavity, and vice versa if a longer time passes.

\begin{figure}
    \includegraphics[width=0.97\linewidth]{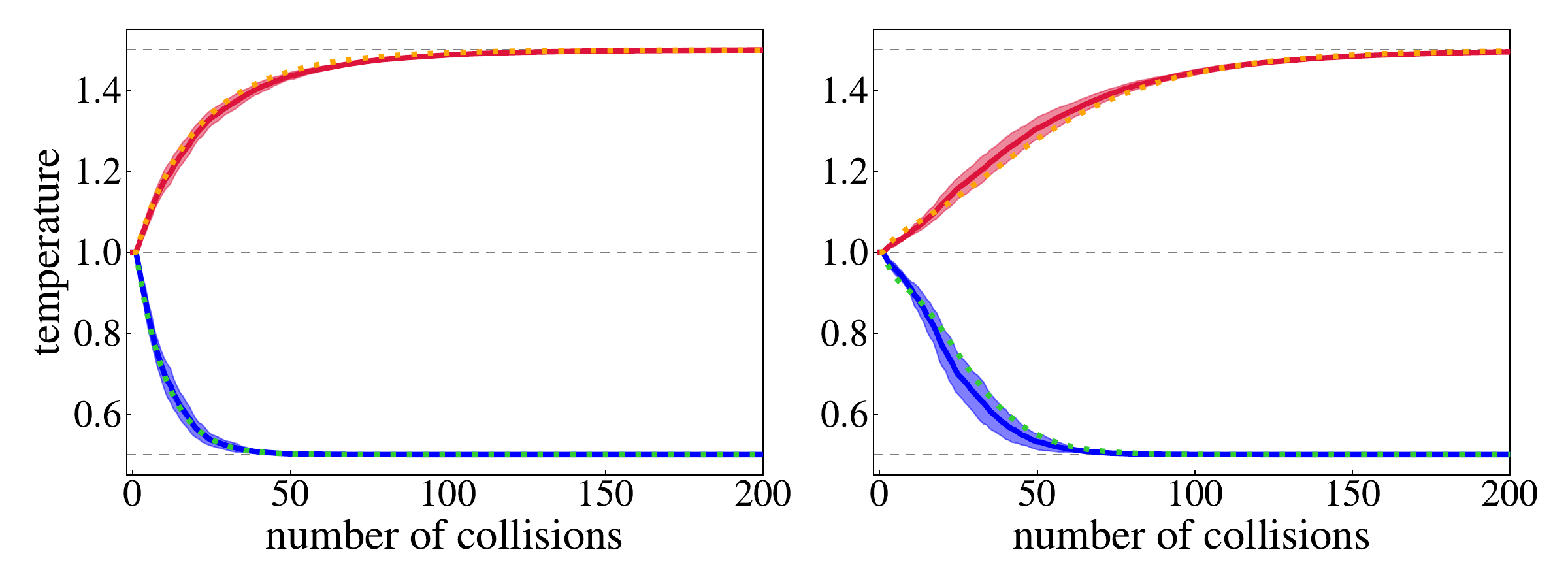}
    \caption{Behaviour of the temperature of the first cavity $S_1$ (left panel) and second cavity $S_2$ (right panel) using interaction times extracted at every step $k$ from a Gaussian distribution: $\Delta t_k = (0.4 \pm 0.2)$, where the standard deviation is used as the error on the mean. Here we selected $\phi^* = \pi/2$, and for the hot curves we use $\alpha^* = 0.2517517$, corresponding to a stationary temperature $T^*=1.5$, while for the cold curves we have $\alpha^* = 0.451963$, corresponding to a stationary temperature $T^*=0.5$.
    The mean value across ten simulations of the evolving temperature is plotted as a continuous line, surrounded by its standard deviation.
    For reference, the constant interaction time case with $\Delta t_k = 0.4$ is plotted with a dotted line.
    It is evident that both heating and cooling the cavities with noise in the interaction times slows down the thermalization process, although the final stable temperature is reached at some point.
    Interaction times are given in units of $1/\Omega$, while temperature is shown in units of $K_B/(\hbar\omega)$.}
    \label{fig:random-times-temperature}
\end{figure}

\subsection{Robustness of the Model}
Here we address the real-world scenario where our parameters, particularly $\phi$ and $\Delta t$ are affected by stochastic noise, so they are not constant but belong to some stochastic distribution.
In fact, suppose a velocity selector is placed in front of the phaseonium beam.
As illustrated in Figure~\ref{fig:phaseonium}, the time each atom spends in a cavity gives the interaction time $\Delta t$, and so it depends on the selected velocity of phaseonium atoms.
We can consider the selector to be flawed and characterized by an absolute error on selected speeds, and those to be Gaussian distributed.
Interaction times will then follow the same distribution, with a variable $\Delta t \equiv \Delta t_k$ changing at each interaction step $k$.
Similarly, in the creation of phaseonium atoms of Eq.~(\ref{def:ancilla-state}), the coherence phase $\phi$ of each atom can be considered as a stochastic Gaussian variable $\phi\equiv\phi_k$, centred on the target coherence phase $\phi^*$.
We then expect a thermalization process for both cavities that takes them to a stochastic temperature $T^* \pm \Delta T^*$.
We will confront this temperature $T^*$ with the expected temperature $T_{\phi^*}$ given by the mean value of the prepared atom coherences.

Thanks to the Markovianity of our approach, where every interaction is independent of the others and the stroboscopic evolution is calculated step-by-step, it is well suited to tackle the problem of stochastic parameters:
we can always apply Eq.~(\ref{eq:final-lindblad-meq}) to each evolution step $k$, using each time the corresponding $\Delta t_k$ and $\phi_k$.

The effect of the noisy interaction times is not expected to significantly affect the thermalization process, since the final temperature of Eq.~(\ref{def:steady-temperature}) does not depend on this parameter: 
no matter the interaction time, every collision will take the system closer to the final stable temperature.
This can be seen from the simulations in Figure~\ref{fig:random-times-temperature}, which shows two thermalization processes for the two cavities from initial temperature $T=1.0$ to temperatures $T_{\phi^*}=1.5$ and $T_{\phi^*}=0.5$, where $\phi_k$ is fixed to $\phi^*$.
The reference cases with constant $\Delta t_k$ are represented with dotted line.
For the stochastic $\Delta t_k$ cases, we report with a continuous line the mean values across different simulations, surrounded by a halo representing the standard deviation.

\begin{figure}
    \includegraphics[width=0.97\linewidth]{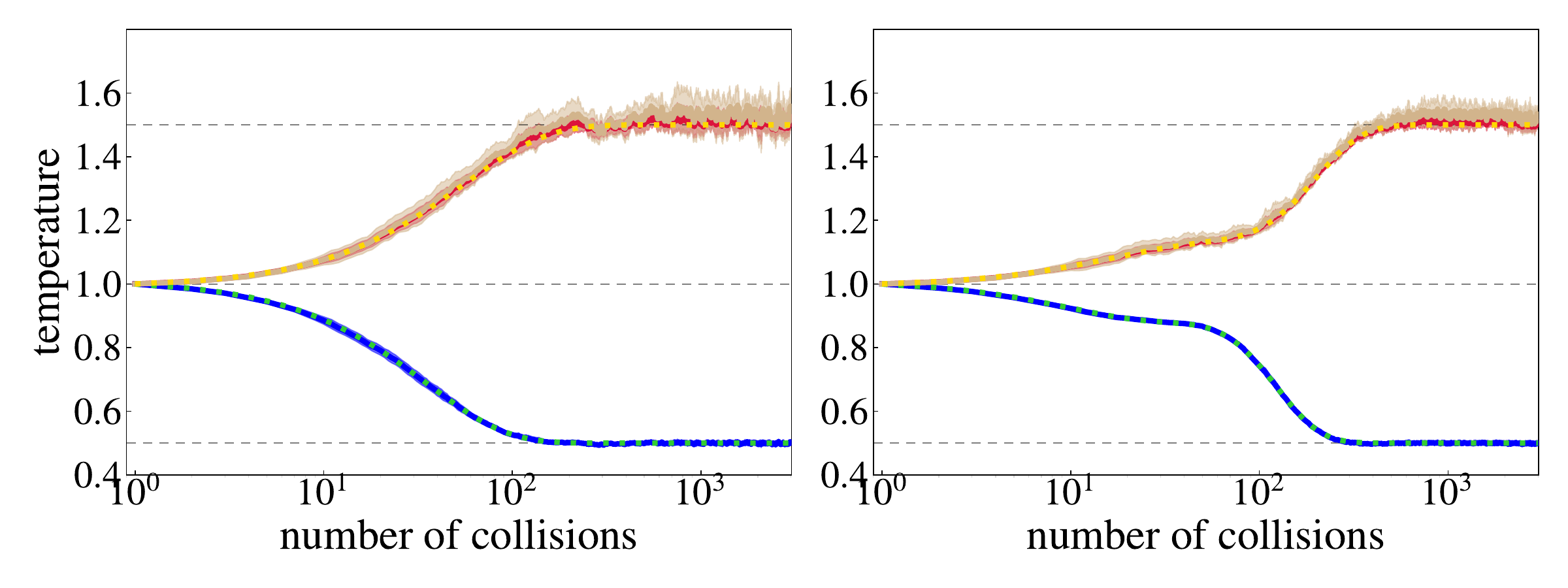}
    \caption{Behaviour of the temperature of the first cavity $S_1$ (left panel) and second cavity $S_2$ (right panel) using coherence phases $\phi_k$ extracted at every step $k$ from a Gaussian distribution with standard deviation $\sigma$. 
    For the cooling process, we select $\alpha^* = 0.25$ and $\phi^*_k = 1.585589386$, with $\sigma=0.2$, corresponding to a stationary temperature $T^*=0.5$. 
    For the heating process, we have two simulations with increasing standard deviation, using $\alpha^* = 0.5$ and $\phi^*_k = 1.26768686$, with $\sigma=0.2$ for the red curve and $\sigma=0.35$ for the brown curve, both corresponding to a stationary temperature $T^*=1.5$.
    Interaction time $\Delta t_k = 0.2$ is fixed.
    The mean value across ten simulations of the evolving temperature is plotted as a continuous line, surrounded by its standard deviation across different simulations.
    For reference, the constant coherence case with $\sigma=0$ is plotted with a dotted line.
    The cooling process is not much affected by random variations in the coherence phase: the first cavity reaches a noisy stable state at $T_1 = (0.498 \pm 0.006)$, and the second thermalizes at $T_2 = (0.498 \pm 0.004)$.
    Noise is higher in the heating scenario, with $T_1 = (1.50\pm0.03)$ and $T_2 = (1.50\pm0.02)$ for the red curve with $\sigma=0.2$ and $T_1 = (1.52\pm0.05)$ and $T_2 = (1.52\pm0.03)$ for the brown curve with $\sigma=0.35$.
    Interaction times are given in units of $1/\Omega$, while temperature is shown in units of $K_B/(\hbar\omega)$.
    }
    \label{fig:random-phases-temperature-standard}
\end{figure}

\begin{figure}
    \includegraphics[width=0.97\linewidth]{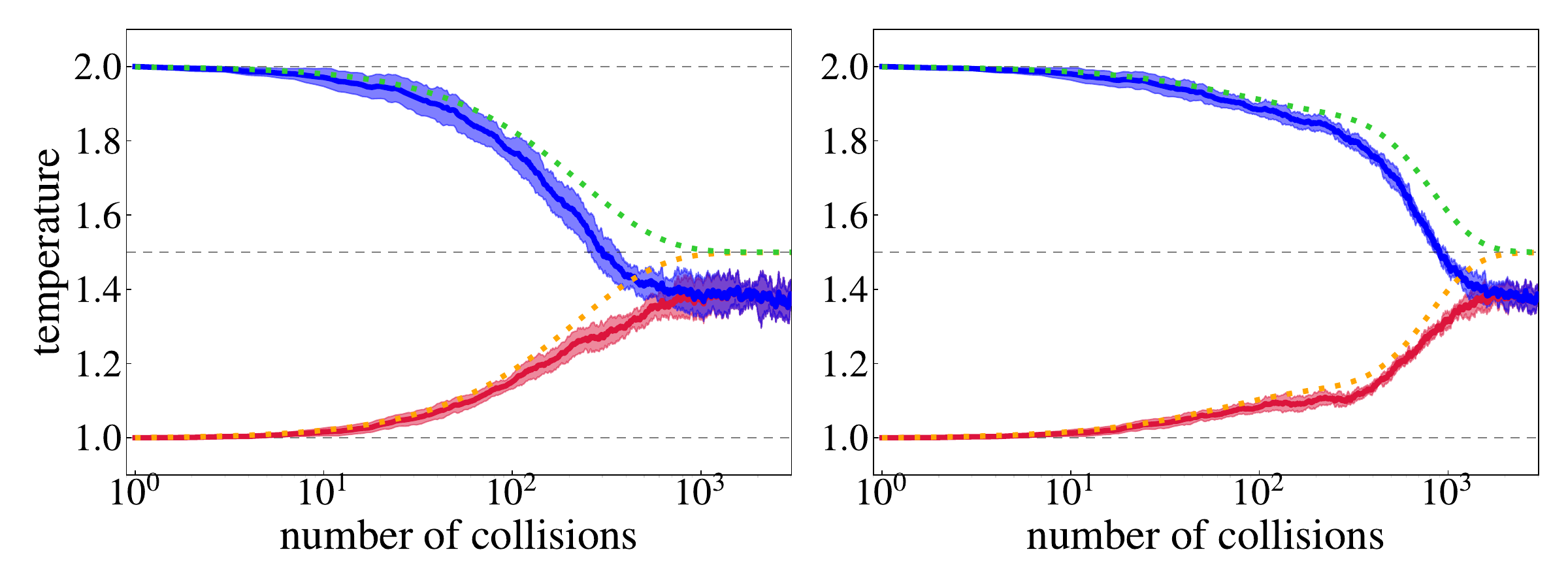}
    \caption{Behaviour of the temperature of the first cavity $S_1$ (right panel) and second cavity $S_2$ (left panel) using coherence phases $\phi_k$ extracted at every step $k$ from a Gaussian distribution with standard deviation $\sigma = 0.2$. 
    Here we selected the same set of parameters $\alpha^* = 0.25$ and $\phi^*_k = (2.404315987 \pm 0.2)$, corresponding to a stationary temperature $T^*=1.5$, but with two different initial conditions for the cavities: starting at temperatures $T_0=2.0$ or $T_0=1.0$.
    Interaction time $\Delta t_k = 0.2$ is fixed.
    The mean value across ten simulations of the evolving temperature is plotted as a continuous line, surrounded by its standard deviation.
    For reference, the constant coherence case with $\phi_k = 2.404315987$ and $\phi_k = 1.585589386$ is plotted with a dotted line.
    A clear difference arises in the reference thermalization at $T^*=1.5$ and the two stochastic processes.
    In fact, in both cases, the first cavity thermalizes to a noisy temperature $T_1=(1.37\pm0.04)$ and the second to $T_2=(1.37\pm0.03)$.
    Interaction times are given in units of $1/\Omega$, while temperature is shown in units of $K_B/(\hbar\omega)$.
    }
    \label{fig:random-phases-temperature-unexpected}
\end{figure}

Different is the case with stochastic coherence phases $\phi_k$ and fixed interaction time $\Delta t_k = \Delta t$, shown in Figure~\ref{fig:random-phases-temperature-standard} and Figure~\ref{fig:random-phases-temperature-unexpected} in the same fashion as the previous figure.
In Figure~\ref{fig:random-phases-temperature-standard} we can see that cooling to temperature $T_{\phi^*}=0.5$ presents only little noise in the final stable temperature, while the heating scenario suffers more, showing manifest noise around the mean value but still getting to an average steady temperature close to the predicted $T_{\phi^*}=1.5$.
Moreover, as can be seen from the two heating processes plotted, the precision of heating is affected by the precision in preparing phaseonium atoms.
Figure~\ref{fig:random-phases-temperature-unexpected} shows different and non-trivial behaviour: heating and cooling with stochastic coherence phases drawn from a different region of the parameter landscape in Figure~\ref{fig:steady-state-temperatures} lead to a steady-state temperature $T^*$ for both cavities which is lower than the expected temperature $T_{\phi^*}$.
This can be understood by looking at the resulting stochastic distribution of ancillas' apparent temperatures defined in Eq.~(\ref{def:apparent-ancillas-temperature}).
In fact, given the nonlinear behaviour of the temperature in $\phi$, the resulting distribution will be right-skewed.
This will shift the mode to the left of the mean, so that we are really sending through the cavities a bigger number of ``cooler'' ancillas with temperature $T_k < T_{\phi^*}$, lowering the final temperature of the systems.

\section{Discussion}\label{sec:discussion}
In this paper, we have studied the dynamics of two optical cavities immersed in a phaseonium bath, highlighting how the quantum coherence present in the initial state of this peculiar atomic environment affects the cavity's evolved state.
In particular, the two-cavity system, assembled in a cascade configuration, thermalizes at a temperature $T_\phi$ which depends on the coherence between the ground levels of the phaseonium.
We have found the quantum dynamical map (see Eq.~(\ref{eq:final-lindblad-meq})) for finite interaction times using only a few assumptions required by collision models.
In general, this equation can be used to calculate the evolution of observables on the system via Eq.~(\ref{eq:final-op-meq}). 

It is also possible to see the unidirectionality of the cascade: the first cavity evolves independently of the second cavity, following the dynamical map of Eq.~(\ref{eq:finitediff_ME}), and it is the first to reach the steady state given by Eq.~(\ref{def:gibbs-state}).
From this moment on, every subsequent ancilla is altered in the same way by the previous interaction and ends up in the state of Eq.~(\ref{eq:steady-ancilla}).
This state then makes the second cavity to thermalize at the same temperature $T_\phi$ of the first cavity. 

Remarkably, we have demonstrated that, by suitably harnessing the excited-state population and coherence phase of each phaseonium atom, the cavity fields can be heated up or cooled down with respect to their initial temperature. 

From an experimental point of view, the use of two harmonic oscillators (here represented by single-mode cavities) and Gaussian states allows us to work with general-purpose systems that are easy to implement in various ways.  
Preparation of phaseonium atoms is also possible via stimulated Raman adiabatic passage (STIRAP) \cite{STIRAP1, STIRAP2}, or similarly fractional STIRAP \cite{fSTIRAP} or fractional Stark-chirped rapid adiabatic passage (f-SCRAP) \cite{fSCRAP}, Morris-Shore transformation \cite{MorrisShoreTransformation}, or quantum Householder reflection \cite{QHR2}. 
The thermodynamic cost of creating phaseonium atoms is studied in \cite{SCULLY2002}.

We have also shown how the interaction time $\Delta t$ can be tuned to achieve faster or slower thermalization.
Besides the velocity selector, another well-known method for modulating the interaction between an atom and a cavity is the Stark shift \cite{raimond_manipulating_2001, davidovich_teleportation_1994, lo_franco_single-shot_2006}.
While phaseonium atoms with level spacing $\hbar\omega'$ enter the cavities with a certain velocity, an external electric field can be applied inside the cavity for a time $\Delta t$ to shift the atom frequency from $\omega'$ to $\omega$ and obtain the required resonant interaction for the desired time.

Also, from the given dynamical equations, it is straightforward to take the continuous limit for $\Delta t \rightarrow 0$, but we emphasize that in real implementations the interaction time is finite and can be used as control parameters. The time spent by an ancilla inside the cavity can be easily set by a velocity selector for phaseonium atoms. 

These results open the way to various prospects. Future works will investigate the role of intra-system quantum correlations, created by the ancillas interacting with both cavities.
Although the cavities are not directly coupled with each other, a unidirectional flow of information is mediated by the ancilla atoms, going from the first cavity to the second one.
Working with quadratic operators as we did in Appendix~\ref{apx:continuous-time}, approximating the time-evolution operator for short collision times, one can exploit the theory of continuous systems' covariance matrices as a useful tool for looking at the evolution of the von Neumann entropy, exchanged mutual information~\cite{mutual_info_Zurek1983, mutual_information_Henderson2001}, quantum discord~\cite{intro_discord_zurek2002, discord1_Ollivier2001, discord2_Henderson2001, discord_for_noncommutativity_Luo2008}, and even entanglement via logarithmic negativity as an entanglement monotone~\cite{ppt_criterion_peres1996, ppt_criterion_horodecki1996, entanglement_monotone_plenio2005}.
The starting point for this further study is given in Appendix~\ref{apx:operators-evolution}.

A further addition to our system will be the implementation of cavity losses to a thermal environment, adding a dissipative term to the master equation to see the effects on the thermalization. This problem is already tackled in Ref.~\cite{saharyan_propagating_2023}, and some specific dissipation models for $\Lambda$ systems in a cavity are considered in Ref.~\cite{obada_moving_2017, chen_exact_2019}. Numerical results for the more simple micromaser are also reported to describe the steady state of the system \cite{xu_influence_2006} and to study the effects on the performances of a quantum micromaser battery \cite{shaghaghi_lossy_2023}. 

Finally, another promising direction to be analysed is the utilization of different types of ancilla phaseonium atoms, prepared with different coherences, to induce a thermodynamic cycle on the cavities.

\begin{acknowledgments}
R.L.F. acknowledges support from European Union -- NextGenerationEU -- grant MUR D.M. 737/2021 -- research project ``IRISQ''. G.M.P. and S.L. acknowledge support by MUR under PRIN Project No. 2022FEXLYB - 
Quantum Reservoir Computing (QuReCo).\\
\end{acknowledgments}

\begin{appendix}

\section{Time-Evolution Operator Expansion}\label{apx:time-evolution-expansion}
This appendix contains the key passages to recover a manageable expression of the evolution operator $\opU$ that rules the evolution of the system $\rho_{k-1}\otimes\eta_k$ and will lead us to the Kraus map \ref{def:kraus1}.
This operator is naturally given by the exponential of the cavity-ancilla interaction \ref{def:interaction}, $\opU = \exp{-i\Delta t\hat V_k}$. 
We can collect the interaction strength $\hbar\Omega$ outside the interaction Hamiltonian and call $\hbar\Omega\Delta t$ as the accumulated Rabi phase $\theta$ during the interaction.
Therefore, we can expand the unitary evolution operator $\opU = exp(i\theta\hat{V}_k)$ as
\begin{equation}\label{eq:opU-series-expansion}
    e^{i\theta V_k} = \sum_{j=0}^\infty \frac{(i\theta)^j}{j!}V_k^j 
    = \mathbb{I}+\sum_{j=1}^\infty \frac{(i\theta)^{2j}}{(2j)!}V_k^{2j}
    + \sum_{j=1}^\infty \frac{(i\theta)^{2j-1}}{(2j-1)!}V_k^{2j-1}
\end{equation}
We will see shortly why this  decomposition in even and odd powers is convenient, but first we must reframe the interaction Hamiltonian in the ancilla basis $\ket{e}$, $\ket{g_1}$, $\ket{g_2}$, that is
\begin{equation}
    \hat{V}_k = \begin{pmatrix}
            0 & \opa & \opa \\
            \opadag & 0 & 0 \\
            \opadag & 0 & 0 
          \end{pmatrix}\,.
\end{equation}
By brute matrix multiplication, we can calculate $\hat{V}_k^2$ and $\hat{V}_k^3$, and easily infer a rule for even and odd powers, as we will show later.
Now, the summation in Eq.~(\ref{eq:opU-series-expansion}) is to be intended matrix element by matrix element.

Referring each element of $\opU$ in row $r$ and columns $l$ as $\opU^{(rl)}$, we can sum up again each series and recover the three by three time-evolution operator $\opU$ given in Eq.~(\ref{def:time-evolution-matrix}).
\begin{widetext}
One finds that the rules for even and odd powers of $\hat{V}_k$ are:
\begin{eqnarray}
    &V^{2j}_k = \begin{pmatrix}
                 2^i(\opa\opadag)^j & 0 & 0 \\
                 0 & 2^{j-1}(\opadag\opa)^j & 2^{j-1}(\opadag\opa)^j \\
                 0 & 2^{j-1}(\opadag\opa)^j & 2^{j-1}(\opadag\opa)^j \\
             \end{pmatrix}\, , \\
    &V^{2j-1}_k = \begin{pmatrix}
                 0 & 2^{j-1}(\opa\opadag)^{j-1}\opa & 2^{j-1}(\opa\opadag)^{j-1}\opa \\
                 2^{j-1}(\opadag\opa)^{j-1}\opadag & 0 & 0 \\
                 2^{j-1}(\opadag\opa)^{j-1}\opadag & 0 & 0 \\
             \end{pmatrix}\, .
\end{eqnarray}
For even powers of the time-evolution operator expansion, we have the elements:
\begin{align}
    \opU^{(11)}=&\sum_{j=1}^\infty \frac{(i\theta)^{2j}}{(2j)!}2^k(\opa\opadag)^j
	= \sum_{j=1}^\infty (-1)^j\frac{\theta^{2j}(\sqrt{2\opa\opadag})^{2j}}{(2j)!}
	= \cos(\theta\sqrt{2\opa\opadag})-\mathbb{I}\;,
    \\
    \opU^{(22)}=\opU^{(23)}=\opU^{(32)}=\opU^{(33)}=
    &\sum_{j=1}^\infty \frac{(i\theta)^{2j}}{(2j)!}2^{j-1}(\opadag\opa)^j
    = \frac{1}{2}\sum_{j=1}^\infty (-1)^j\frac{\theta^{2j}(\sqrt{2\opadag\opa})^{2j}}{(2j)!}
    = \frac{1}{2} \left[\cos(\theta\sqrt{2\opadag\opa}) - \mathbb{I}\right]
\;,\end{align}
and for odd powers we have the elements:
\begin{align}
    \opU^{(12)}=\opU^{(13)}=&
    \sum_{j=1}^\infty \frac{(i\theta)^{2j-1}}{(2j-1)!}2^{j-1}\opadag(\opa\opadag)^{j-1}
    = \sum_{j=1}^\infty(-1)^j\opadag\frac{\theta^{2j-1}(\sqrt{2\opa\opadag})^{jk-1}}{(2j-1)!\sqrt{2\opa\opadag}}
    = i\opadag\frac{\sin(\theta\sqrt{2\opa\opadag})}{\sqrt{2\opa\opadag}} \;,
    \\
    \opU^{(21)}=\opU^{(31)}=&\sum_{k=1}^\infty \frac{(i\theta)^{2j-1}}{(2j-1)!}2^{j-1}(\opa\opadag)^{k-1}\opa 
    = i \sum_{j=1}^\infty  (-1)^j\frac{\theta^{2j-1}(\sqrt{2\opa\opadag})^{2j-1}}{(2j-1)!\sqrt{2\opa\opadag}}\opa
    = i\frac{\sin(\theta\sqrt{2\opa\opadag})}{\sqrt{2\opa\opadag}}\opa
\;,\end{align}
having used in the first line the equivalence $f(\opadag\opa)\opadag = \opadag f(\opa\opadag)$. 
\end{widetext}

We can thus define the  photon operators that appear in the elements of the time-evolution operator as in \cref{def:photonic-ops},
and with those we can rewrite the exponential operator in terms of the $3\times3$ matrix representation in the ancilla's basis
\begin{equation}
    \hat{U}^\dagger_k = e^{-i\theta \hat{V}_k} =
    \begin{pmatrix}
        \hat{C} & -i S^\dagger & -i S^\dagger \\
        -i \hat{S} &  \tfrac{1}{2}(\hat{C}' +\mathbb{I}) & \tfrac{1}{2}(\hat{C}' -\mathbb{I}) \\
        -i \hat{S} &  \tfrac{1}{2}(\hat{C}' -\mathbb{I}) & \tfrac{1}{2}(\hat{C}' +\mathbb{I})
    \end{pmatrix}
\;.\end{equation}

\begin{equation}
    \hat{U}_k = e^{i\theta \hat{V}_k} =
    \begin{pmatrix}
        \hat{C} & i S^\dagger & i S^\dagger \\
        i \hat{S} &  \tfrac{1}{2}(\hat{C}' +\mathbb{I}) & \tfrac{1}{2}(\hat{C}' -\mathbb{I}) \\
        i \hat{S} &  \tfrac{1}{2}(\hat{C}' -\mathbb{I}) & \tfrac{1}{2}(\hat{C}' +\mathbb{I})
    \end{pmatrix}
\;.\end{equation}

Note that now the time-dependence of the evolution is hidden inside the sinusoidal functions of photon operators $\hat{C}$, $\hat{C}'$ and $\hat{S}$.

\section{Continuous Time Limit}\label{apx:continuous-time}

\renewcommand{\a}{\hat{a}}
\newcommand{\ad}{\hat{a}^\dagger}
\renewcommand{\b}{\hat{b}}
\newcommand{\bd}{\hat{b}^\dagger}
\renewcommand{\tt}{\theta^2}

We show here the continuous time limit of the master equation \ref{eq:finitediff_ME} for the cascade system.
Revising Ref.~\cite{ciccarello2021}, continuous time is a standard approximation in collision models approaches, where system-ancilla interactions are taken to be short enough to allow a second-order approximation of the time-evolution operator $\opU$.
As long as collisions \emph{must happen}, the limit $\Delta t \rightarrow 0$ can survive only for diverging interaction strengths; 
else, one focus on evolution times much larger than $\Delta t$, so that the stroboscopic dynamics happening at discrete time steps $t_n = n\Delta t$ can be replaced with a continuous time variable $t$, and finite differences can be replaced with differentials.
This is called \emph{coarse graining}.
In this approximation, photonic operators in \cref{def:photonic-ops} become
\begin{equation}
    C=\mathbb{I} - \theta^2aa^\dagger \qquad
    C'= \mathbb{I} - \theta^2a^\dagger a \qquad
    S = \theta a^\dagger 
\end{equation}
and denoting $\a$ ($\ad$) and $\b$ ($\bd$) as the annihilation (creation) operators for the first  and second cavity respectively, the Kraus operators in \cref{def:kraus2} can be written as
\begin{align}
    E_0 &= \sqrt{1-\frac{\gamma_\alpha}{2}-\frac{\gamma_\beta}{2}}\mathbb{I}\, , 
    \\[1em]
    E_1 &= \sqrt{\frac{\gamma_\alpha}{2}}\left[ 1-\tt\b\bd-\tt\a\ad-2\tt\ad\b \right]\, ,
    \\[1em]
    E_2 &= \sqrt{\gamma_\alpha}\left[ \theta\ad + \theta\bd \right]\, ,
    \\[1em]
    E_3 &= \sqrt{\gamma_\beta}\left[ \theta\a + \theta\b \right]\, ,
    \\[1em]
    E_4 &= \sqrt{\frac{\gamma_\beta}{2}}\left[ 1-\tt\bd\b-\tt\ad\a-2\tt\a\bd \right]\, ,
\end{align}
We can now use these expressions to rewrite the discrete master equation
(\ref{eq:finitediff_ME}) as

\begin{equation}
\label{def:lindbladian-evolution}
    \frac{d\rho(t)}{d t}
    =-i[\hat{H}_\text{eff},\rho]+ \gamma'_\alpha \,\mathcal{D}[\hat{a}^\dagger{+}\hat{b}^\dagger]\,\hat{\rho}(t)+\gamma'_\beta \,\mathcal{D}[\hat{a}{+}\hat{b}]\,\hat{\rho}(t) 
\end{equation}

where $\hat{H}_\text{eff}=i(\gamma'_\beta-\gamma'_\alpha)(\ad \b-\bd\a)/2$ and $\gamma'_{\alpha,\beta}=\gamma_{\alpha,\beta}\Omega^2\Delta t$, considering always $\Omega$ such that $\Omega^2 \Delta t$ converges in the continuous limit.
This cascade master equation is comparable to that obtained by the standard collision model methods explained in \cite{ciccarello2021}.
\section{Cavity Steady State}\label{apx:steady-state}
Here we look for the steady state of one cavity after a {\it sufficient} number of collisions. 
The stationary condition reads
\begin{equation}
   \Delta\rho^*= \sum_i \hat E_i\rho^*\hat E_i^\dagger-\rho^*=0 \,.
\end{equation}
with Kraus operators given in Eq.~(\ref{def:kraus1}).
This must be true for each expectation value in the system's basis, so using the relations $f(\opa\opadag)\ket{n}=f(n+1)\ket{n}$ and $f(\opadag\opa)\ket{n}=f(n)\ket{n}$ we have that off-diagonal elements are null,
\begin{equation}
\begin{split}
    \Delta{\rho^*}^{n}_m \equiv \mel**{n}{\Delta\rho^*}{m} =  0& \;, \\
    n\neq m
\end{split}
\end{equation}
and for $n=m$ we report the relative equation
\begin{widetext}
\begin{eqnarray}
    \left[
    \frac{\gamma_\alpha}{2}\sin^2{\theta\sqrt{2n}}\right]\,{\rho^*}^{n-1}_{n-1}+
    \left[\frac{\gamma_\alpha}{2}\cos^2{\theta\sqrt{2(n{+}1)}}+\frac{\gamma_\beta}{2}\cos^2{\theta\sqrt{2n}}+\frac{\beta^2}{2}(1-\cos\phi)\right]\,{\rho^*}^{n}_n+
    \left[\frac{\gamma_\beta}{2}\sin^2{\theta\sqrt{2(n{+}1)}}\right]\,{\rho^*}^{n+1}_{n+1}
    &={\rho^*}^{n}_n
\end{eqnarray}
\end{widetext}
This can be recursively resolved obtaining
\begin{equation}
    {\rho^*}^{n}_n=\frac{\gamma_\alpha}{\gamma_\beta}{\rho^*}^{n-1}_{n-1}=\left(\frac{\gamma_\alpha}{\gamma_\beta}\right)^n{\rho^*}^{0}_0 \; .
\end{equation}

We can now refer to the element ${\rho^*}^{0}_{0}$ as $1/Z$ and $\left(\gamma_\alpha / \gamma_\beta \right)^n~=~\exp(-n\hbar\omega / K_BT_\phi)$ to write the steady state of the system as a Gibbs State:
\begin{equation}\label{gibbs}
    \rho^* = \sum_n^\infty\frac{\exp(-E_n/K_BT_\phi)}{Z}\dyad{n} \,.
\end{equation}
and thus define the steady state phase-dependent temperature $T_\phi$ given in \ref{def:steady-temperature}.

As the first cavity thermalizes to this temperature, ancillas will cease to interact with it and the second cavity will start to see the originally prepared ancillas bearing the same $T_\phi$ and to evolve like it was a single cavity, finally thermalizing to $T_\phi$.

Since this derivation of the steady state is the most general, it will hold as for the coarse-grained dynamics expressed by Eq.~(\ref{def:lindbladian-evolution}). The steady-state temperature, in fact, does not depend on the interaction time $\Delta t$.

\section{Ancillas Decoherence}\label{apx:decoherence}
Here we study how our results would be modified if we add a decoherence process to ancillas.
In fact, we can suppose ancillas to interact with the environment before and after entering the cavities. 
We can think of the decoherence process $U_{d}(t)$ as a process which takes the coherent ancillas towards their thermal counterpart after some time $\tau$:
\begin{equation}
    \eta = \begin{pmatrix}
        a & 0 & 0 \\
        0 & \frac{b}{2} & c \\
        0 & c^* & \frac{b}{2}
    \end{pmatrix} \longrightarrow 
    U_d(\tau)\eta U_d^\dagger(\tau) =
    \begin{pmatrix}
        a & 0 & 0 \\
        0 & \frac{b}{2} & 0 \\
        0 & 0 & \frac{b}{2}
    \end{pmatrix} \,.
\end{equation}
Note that with this notation we define dissipation rates as $\gamma_\alpha=2a$ and $\gamma_\beta=b + (c+c^*) = b + \mathcal{R}e(c)$, where $\mathcal{R}e(c)$ is the real part of the coherences.
We can thus model the process $U_d(t)$ simply multiplying the coherences with a time-dependent coefficient $\epsilon(t)$ which goes from $1$ to $0$ as $t$ goes from $0$ to $\tau$, so that the coherences themselves become time-dependent: $c(t) = \epsilon(t)c$.
Moreover, we can suppose that if the first cavity is close enough to the phaseonium source, the effect of the decoherence on ancillas seen by $S_1$ is negligible.
This cavity will thus thermalize to temperature $K_BT^*_1$.
For a given ``time of flight'' $\bar{t}$ of ancillas between one cavity and the other, the decohered ancilla seen from the second cavity will have the form
\begin{equation}
    \begin{pmatrix}
        a & 0 & 0 \\
        0 & \frac{b}{2} & \epsilon(\bar{t})c \\
        0 & \epsilon(\bar{t})c^* & \frac{b}{2}
    \end{pmatrix} \,.
\end{equation}
Accordingly, only the dissipation rate $\gamma_\beta$ of the second cavity will be modified.
This cavity will thus thermalize to a steady state with temperature different from the first cavity:
\begin{align}
    &K_BT^*_1 = -\frac{\hbar\omega}{\ln{(2a/(b+\mathcal{R}e\{c\})}}\,, \\
    &K_BT^*_2 = -\frac{\hbar\omega}{\ln{(2a/(b+\epsilon(\bar{t})\mathcal{R}e\{c\})}}\,, \\[5pt]
    &\frac{\hbar\omega}{T^*_1} - \frac{\hbar\omega}{T^*_2} = \ln{\left(\frac{b+\mathcal{R}e\{c\}}{b+\epsilon(\bar{t})\mathcal{R}e\{c\}}\right)}\,.
\end{align}
or 
\begin{equation}
    \frac{K_BT^*_2}{\hbar\omega} = K_BT^*_1 \bigg/ \left( \hbar\omega - K_BT^*_1\ln{\left(\frac{b+\mathcal{R}e\{c\}}{b+\epsilon(\bar{t})\mathcal{R}e\{c\}}\right)} \right)\;.
\end{equation}

\section{Operators Evolution}\label{apx:operators-evolution}
The variation of the expectation value of an observable acting on the Hilbert space of the system in each collision is given by, (cf. Eq~(\ref{eq:finitediff_ME}):
\begin{equation}
 \Tr{\Obs\,\Delta\rho_{k}}=\Tr{\Obs\left(\sum_{i=0}^{4} \hat E_i \rho_k \hat E_i^\dagger-\rho_k\right)}
\end{equation}
Thanks to the permutation property of the trace, this can be recast into
\begin{equation}
    \Tr{\left(\sum_{i=0}^{4} \hat E_i^\dagger\Obs\hat E_i -\Obs\right)\rho_k}=\Tr{\Delta \Obs\, \rho_k}
\end{equation}
leading to the following discrete master equation for $\ave{\Obs}_k$
\begin{equation}
    \Delta\ave{\Obs}_k= \sum_{i=1}^4\ave{\widetilde{\mathcal{D}}[E_i]\Obs}_k 
    \label{eq:final-op-meq}
\end{equation}
where $\widetilde{\mathcal{D}}[\hat E]\Obs=\hat E^\dagger\Obs\hat E-1/2\{\hat E^\dagger\hat E,\Obs\}$.

As a figure of merit, we can give the expression for the time evolution of the average number of photons inside one cavity $\Delta \ave{n}_k$,
\begin{equation}\label{eq:n-meq-apx}
\Delta \ave{n}_k=\sum_n \rho^{n,n}_k \left[\frac{\gamma_\alpha}{2}\sin^2\left(\theta\sqrt{2(n+1)}\right)-\frac{\gamma_\beta}{2}\sin^2\left(\theta\sqrt{2n}\right)\right]
\end{equation}
This dictates the rate of heating (if $\Delta \ave{n}_k$ is positive) or cooling (if $\Delta \ave{n}_k$ is negative) of one cavity, and it is dependent on $\cos\phi$ inside the $\gamma_\beta$ coefficient and on $\theta = \Omega\Delta t$ via two squared sinusoidal functions.

Taking the continuous time limit as in Eq.~(\ref{def:lindbladian-evolution}), we obtain
\begin{align}\label{def:Ops-lindbladian-evolution-apx}
    \frac{d\ave{\Obs}}{d t}
    = \ave{i[\hat{H}_\text{eff},\rho]} +  \gamma'_\alpha \,\ave{\widetilde{\mathcal{D}}[\hat{a}^\dagger{+}\hat{b}^\dagger]\,\Obs}+\gamma'_\beta \,\ave{\widetilde{\mathcal{D}}[\hat{a}{+}\hat{b}]\,\Obs}
\end{align} 
We emphasize a crucial distinction from Eq.~(\ref{eq:final-op-meq}): the presented master equation is quadratic in the cavity operators, thereby preserving Gaussianity. Consequently, the utility of working with Gaussian states becomes evident. Gaussian states, when subjected to such a coarse-grained quantum map, undergo evolution while retaining their Gaussian nature.
\end{appendix}

\end{document}